%% file: local_sensing_rev2.tex
\documentclass[%
reprint,
superscriptaddress,
showpacs,preprintnumbers,
 amsmath,amssymb,
 aps,
pra,
]{revtex4-1}

\usepackage{graphicx}   
\usepackage{dcolumn}
\usepackage{bm}
\usepackage{hyperref}

\usepackage{epstopdf}
\usepackage{braket}
\usepackage{siunitx}

\binoppenalty=\maxdimen
\relpenalty=\maxdimen

\begin{document}


\title{Local Sensing with the Multi-Level AC Stark Effect}

\author{Andre Schneider}	\affiliation{Institute of Physics, Karlsruhe Institute of Technology, 76131 Karlsruhe, Germany}
\email{andre.schneider@kit.edu}
\author{Jochen Braum\"uller}\affiliation{Institute of Physics, Karlsruhe Institute of Technology, 76131 Karlsruhe, Germany}
\author{Lingzhen Guo}		\affiliation{Institute for Theoretical Solid State Physics, Karlsruhe Institute of Technology, 76131 Karlsruhe, Germany}
\author{Patrizia Stehle}	\affiliation{Institute of Physics, Karlsruhe Institute of Technology, 76131 Karlsruhe, Germany}
\author{Hannes Rotzinger}	\affiliation{Institute of Physics, Karlsruhe Institute of Technology, 76131 Karlsruhe, Germany}
\author{Michael Marthaler}
	\affiliation{Institute for Theoretical Solid State Physics, Karlsruhe Institute of Technology, 76131 Karlsruhe, Germany}
	\affiliation{Theoretische Physik, Universit\"at des Saarlandes, 66123 Saarbr\"ucken, Germany}
\author{Alexey V. Ustinov}	
	\affiliation{Institute of Physics, Karlsruhe Institute of Technology, 76131 Karlsruhe, Germany}
	\affiliation{Russian Quantum Center, National University of Science and Technology MISIS, 119049 Moscow, Russia}
\author{Martin Weides}		
	\affiliation{Institute of Physics, Karlsruhe Institute of Technology, 76131 Karlsruhe, Germany}
	\affiliation{Institute of Physics, Johannes Gutenberg University Mainz, 55099 Mainz, Germany}
	\affiliation{Materials Science in Mainz, University Mainz, 55099 Mainz, Germany}
	\affiliation{School of Engineering, University of Glasgow, Glasgow G12 8QQ, UK}

\date{\today}


\begin{abstract}

Analyzing weak microwave signals in the GHz regime is a challenging task if the signal level is very low and the photon energy widely undefined.
A superconducting qubit can detect signals in the low photon regime, but due to its discrete level structure, it is only sensitive to photons of certain energies.
With a multi-level quantum system (qudit) in contrast, the unknown signal frequency and amplitude can be deduced from the higher level AC Stark shift. 
The measurement accuracy is given by the signal amplitude, its detuning from the discrete qudit energy level structure and the anharmonicity. We demonstrate an energy sensitivity in the order of $10^{-3}$ with a measurement range of more than $\SI{1}{\giga\hertz}$.
Here, using a transmon qubit, we experimentally observe shifts in the transition frequencies involving up to three excited levels. 
These shifts are in good agreement with an analytic circuit model and master equation simulations.
For large detunings, we find the shifts to
scale linearly with the power of the applied microwave drive.
Exploiting the effect, we demonstrated a power meter which makes it possible to characterize the microwave transmission from source to sample.
\end{abstract}


\maketitle

\section{\label{sec:introduction}{Introduction}}
While early experiments in the field of superconducting circuit quantum electrodynamics (cQED) only included very few excitations  \cite{Wallraff2004}, 
recent studies start exploring highly populated resonant circuits \cite{Wang_Science16}. 
Large photon numbers become more and more relevant for quantum computing and simulation protocols, e.g.\@ by enabling
high fidelity state determination \cite{Sank_arXiv,Reed2010}, all-microwave multi-qubit gates \cite{Chow2011,Paik2016} or dynamic coupling schemes \cite{Braumueller2016, BallesterPRX2012}.
Their microwave electromagnetic fields cause level shifts, called AC Stark or AC Zeeman for electric or magnetic coupling to the quantum circuit.
We investigate these shifts of a multilevel anharmonic system under the influence of a drive and find good agreement with the calculated circuit's response.
We can also use an analytic model to deduce the frequency and amplitude of a microwave drive from the observed shift, enabling to use the quantum circuit as a sensor for microwave signals.
Superconducting qubits have already been used as microwave sensors to calibrate the signal power at the chip, but the calibration with these methods is limited to discrete transition frequencies \cite{Baur2009,Astafiev2010,BraumuellerPRB15} or to the sub-GHz regime with tunable qubits \cite{Jerger2017}.

In this paper, we study the response of an anharmonic circuit featuring higher levels. 
Such multi-level quantum elements (so-called qudits) are a possible path in the pursuit of scaling quantum circuits up.
In the literature, the AC Stark shift of the first transition $\ket{0} \leftrightarrow \ket{1}$ for multi-level circuits with weak \cite{Strauch_IEEE2007} and strong \cite{SchusterPRL2005} anharmonicity has been observed and analytically explained. 
We strive to understand the fundamental level structure of qudits driven by increasingly large signal amplitudes. 
Having a detailed model of the qudit behavior, we gain the ability of taking the qudit as a sensor for probing the excitation number states in coupled harmonic systems like microwave resonators or magnonic systems.
Applying a short but strong microwave pulse can also be used to quickly detune even non-tunable qudits in a controlled manner, which can be employed for the realization of quantum gates.

In the following, we derive a theoretical description of the effect using a fourth order perturbation theory on a driven multi-level anharmonic system. We validate the findings experimentally and show good agreement with a master equation simulation of the transmon Hamiltonian. 
In a second experiment, we apply these findings and successfully demonstrate amplitude and frequency sensing of an external microwave drive with our technique.

\section{\label{sec:theory}{Theoretical Analysis}}
We start our analysis with the unit cell of a cQED system, consisting of an anharmonic and a harmonic oscillator, expressed by the generalized Hamiltonian \cite{Koch_TransmonPRA07}
\begin{eqnarray}
\label{eq:general_hamiltonian}
H/\hbar &=& \sum_j \omega_j \ket{j}\bra{j} + \omega_\mathrm{r}b^\dag b + \sum_{i,j} g_{ij} \ket{i}\bra{j}(b+b^\dag)
\end{eqnarray}
where $b^\dag$, $b$ and $\omega_\mathrm{r}$ are the raising/lowering operator and  the resonance frequency of the harmonic system and $\omega_j$ and $g_{ij}$ are the transition frequencies and coupling matrix elements of the anharmonic system. 

In a regime where the anharmonicity is large compared to the bandwidth of all applied signals, we can reduce the anharmonic system to an effective two level system (qubit) in a good approximation. 
Additionally, in the dispersive regime \citep{Wallraff2004,BlaisPRA2004}, where the transition frequencies of both systems are strongly detuned, Eq.~(\ref{eq:general_hamiltonian}) is reduced to the approximately diagonalized Jaynes-Cummings Hamiltonian 
\begin{eqnarray}
\label{Eq:jaynescummings}
H/\hbar &=& \omega_\mathrm{r}' \left(b^{\dag}b+\frac{1}{2}\right)+{\left(\omega_\mathrm{q}^0+\frac{2g^2}{\Delta} b^{\dag}b+\frac{g^2}{\Delta}\right)}\frac{\sigma^z}{2}\,.
\end{eqnarray}
The mixed term $\frac{\,g^2}{\Delta}\,b^{\dag}b\, \sigma^z$ can be interpreted as a Lamb shift of the resonator frequency depending on the qubit state, which is commonly used for read-out of the qubit state. 
Vice versa, it can be seen as a shift of the dressed qubit frequency $\omega_\mathrm{q}$ depending on the resonator population, the so-called AC Stark shift \cite{SchusterPRL2005,Gambetta_PRA06}.
Here, $g$ denotes the coupling strength between qubit and resonator and $\Delta=\omega_\mathrm{r}'-\omega_\mathrm{q}^0$ is the detuning between both subsystems.
As this coupling can either be capacitive or inductive, we can also use the term AC Zeeman shift to describe the observed effect.
The qubit levels shift by $\pm n g^2/ \Delta$, being linear in the photon number $n$, i.e.\@ the power applied to the resonator.

Going back to Eq.~(\ref{eq:general_hamiltonian}),
we now model our anharmonic circuit, commonly known as a transmon \cite{Koch_TransmonPRA07}, by 
\begin{eqnarray}
H_\mathrm{q}/\hbar = \omega_\mathrm{q} a^\dag a - \frac{\gamma}{4}a^\dag a ( a^\dag a +1),
\end{eqnarray}
where $a^\dag$ and $a$ are the anharmonic raising and lowering operators.
The circuit anharmonicity is $-\hbar \frac{\gamma}{2}= -E_\mathrm{C}$.
For our analysis, we add an external drive $H_\mathrm{D}/\hbar = A_\mathrm{D}(a+a^\dag)\cos{\omega_\mathrm{D}t}$ with drive frequency $\omega_{\mathrm{D}}$.
The amplitude $A_\mathrm{D}$ is well controllable and only depends on the drive power $P_\mathrm{D}$.
This drive can also originate from the field of a strongly populated cavity, where the
resonator photon number $n = \braket{b}^2 \propto A_\mathrm{D}^2$  is not linear to the power $P_\mathrm{r}$ sent to the resonator due to its changing quality factor \cite{Wang_APL}. We therefore stick to the externally supplied drive for the rest of this work.

In the following, we drop the coupled readout resonator for simplicity, and rewrite the driven transmon circuit as
\begin{eqnarray}
\label{eq:driven_transmon}
H/\hbar &=& \omega_\mathrm{q} a^\dag a - \frac{\gamma}{4}a^\dag a (a^\dag a +1)
+ A_\mathrm{D} (a+a^\dag)\cos{\omega_\mathrm{D} t}\, 
\end{eqnarray}
where constant energies and corrections linear in $\gamma$ and $a^\dag$ have been dropped for simplicity. 
We only take into account a first-order nonlinearity, which generalizes the following calculations for other anharmonic systems.
The transition from the $\ket{k}$ to the $\ket{k+1}$ state has an energy of 
\begin{equation}
E_{k \rightarrow k+1}/\hbar = \omega_\mathrm{q}-\frac{\gamma}{2}(k+1) .
\end{equation}
Note that the first transition
$E_{0 \rightarrow 1}/\hbar = \omega_\mathrm{\mathrm{q}}-\frac{\gamma}{2}$ is already detuned from the harmonic transition.
Transforming the Hamiltonian to the rotating frame of the drive yields
\begin{eqnarray}
\label{Eq:RWA}
 H_\mathrm{R} /\hbar &=& U^{\dag}(t)H_\mathrm{AC} U(t)/\hbar + \mathrm{i} U^{\dag}(t)\dot{U}(t)\, ,\,\, U(t)=e^{\mathrm{i}\omega_{\mathrm{D}} a^{\dag}a t}\nonumber\\
           &\approx &(\omega_\mathrm{q}-\omega_{\mathrm{D}})a^{\dag}a-\frac{\gamma}{4} a^{\dag}a(a^{\dag} a+1)+\frac{A_\mathrm{D}}{2}(a+a^{\dag})\mathrm{,}
\end{eqnarray}
omitting fast rotating terms in a rotating wave approximation.
To get an expression for the transmon energy levels depending on the drive amplitude $A_\mathrm{D}$, we need to treat the drive as a perturbation. 
However, the last term $\frac{ A_\mathrm{D}}{2}(a+a^\dag)$ is not small compared to the other terms and we have to transform the Hamiltonian by 
a displacement operator $ D_\alpha\equiv e^{\alpha
a^\dagger-\alpha^* a}$, with $ D^\dagger_\alpha
aD_\alpha=a+\alpha$:
\begin{eqnarray}\label{H}
\tilde{H}_\mathrm{R}/\hbar &=&D^\dagger_\alpha H_\mathrm{R} D_\alpha /\hbar = \left(
\omega_\mathrm{q}-\omega_\mathrm{D}-\frac{\gamma}{4}-\gamma|\alpha|^2\right)a^\dagger a\nonumber\\
&&-\frac{\gamma}{4}(a^\dagger a)^2
+({\beta}^*a+
a^\dagger\beta)
-\frac{\gamma}{2}\left(  \alpha^*a^\dag a a+\alpha a^\dag a^\dag a  \right)\nonumber\\
&&-\frac{\gamma}{4}(\alpha^2 a^{\dagger2}+\alpha^{*2}
a^2)+\mathrm{const}
\end{eqnarray}
with introducing the parameter $\beta$ as
\begin{equation}\label{Eq:alphaTilde}
\beta\equiv \alpha\Big[
\omega_\mathrm{q}-\omega_\mathrm{D}-\frac{\gamma}{2}(1 + |\alpha|^2)\Big]
+\frac{A_\mathrm{D}}{2}\,.\end{equation}
We can cancel the linear term 
$({\beta}^*a+
a^\dagger\beta)$ in Eq.~(\ref{H}) by choosing the free parameter $\alpha$ such that 
$\beta=0$. Then the Hamiltonian becomes
\begin{eqnarray}\label{HR}
\tilde{H}_\mathrm{R} /\hbar
&=&\underbrace{\left(  
\omega_\mathrm{q}-\omega_\mathrm{D}-\frac{\gamma}{4}-\gamma|\alpha|^2\right)a^\dagger
a-\frac{\gamma}{4}(a^\dagger a)^2}_{\tilde{H}^0_\mathrm{R}/\hbar}\\
&&\underbrace{-\frac{\gamma}{4}(\alpha^2 a^{\dagger2}+\alpha^{*2} a^2)-\frac{\gamma}{2}\left(  \alpha^*a^\dag a^2+\alpha a^{\dag 2} a  \right)}_{\tilde{H}^1_\mathrm{R}/\hbar}.\nonumber
\end{eqnarray}
We use perturbation theory to obtain the eigenenergies of $\tilde{H}_\mathrm{R}$, where the full calculation up to the fourth order in perturbation theory is given in the supplement. To zeroth order we get:
\begin{eqnarray}\label{Eq:En}
\tilde{E}_k /\hbar &\approx &\left(  
\omega_\mathrm{q}-\omega_\mathrm{D}-\frac{\gamma}{4}-\gamma|\alpha|^2  \right) k-\frac{\gamma}{4}k^2 
+\dots\,.
\end{eqnarray}

From the requirement $\beta=0$ (Eq.~(\ref{Eq:alphaTilde})), we can determine the
parameter $\alpha$ by introducing the detuning $\Delta = \omega_\mathrm{D}-\omega_\mathrm{q}+\frac{\gamma}{2}$ between the drive and the first qudit transition
\begin{equation}
\gamma\alpha|\alpha|^2 + 2  \Delta \alpha -A_\mathrm{D} = 0.
\label{}
\end{equation}
In general there are three solutions for $\alpha$. For the case of $\Delta/\gamma > 0$ which applies for this paper, there is only one real solution:
\begin{equation}
\label{eq:alpha_eq}
\alpha = \frac{{2}^{1/3} \left(\sqrt{81 A_\mathrm{D}^2 \gamma +96 \Delta ^3}+9 A_\mathrm{D} \sqrt{\gamma }\right)^{2/3}-4 \sqrt[3]{3} \Delta }{ \left(  { 36 \left[  {3\gamma ^3 \left(27 A_\mathrm{D}^2 \gamma +32 \Delta ^3\right)} \right]^{1/2}+324 A_\mathrm{D} \gamma ^2}\right)^{1/3}}.
\end{equation}

For $\Delta/\gamma < 0$, a critical drive amplitude exists, below which three real solutions for $\alpha$ can be found. 
In this case, it is more likely to directly excite higher qudit levels with the strong drive as the levels bend towards the drive frequency with stronger driving. For all practical applications, where the qudit is used as a sensor, we are therefore only interested in the case of $\Delta/\gamma >0$.

\begin{figure}[tb]
	\includegraphics[width=\columnwidth]{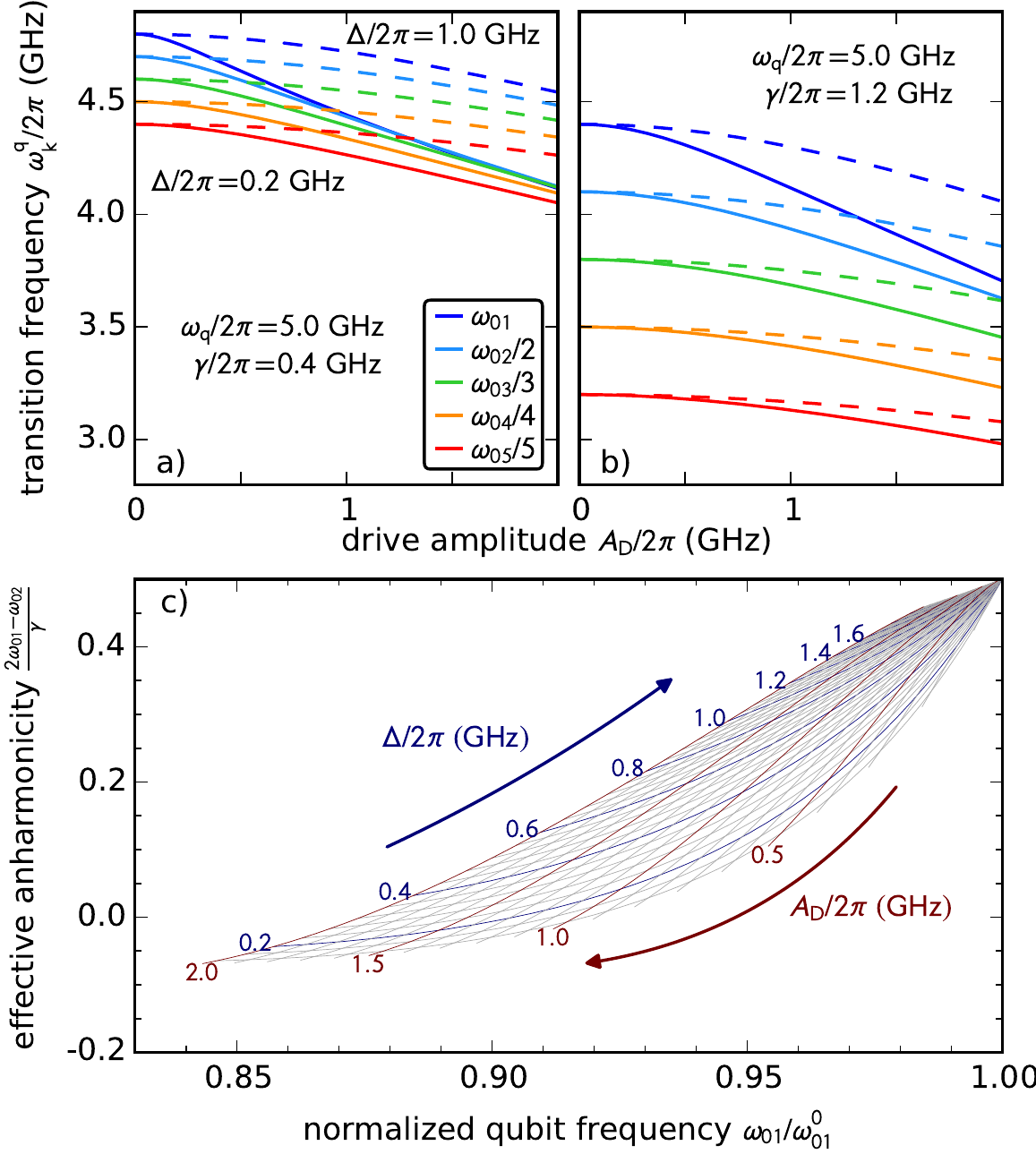}
	\caption{Analytically calculated AC Stark shift for (a) a weakly anharmonic circuit ($\omega_\mathrm{q}/2\pi=5\,\mathrm{GHz}$, $\gamma/2\pi=0.4 \,\mathrm{GHz}$, solid lines) using perturbation theory up to fourth order. For strong driving amplitudes, a crossing of the multi-photon transitions can be observed. The dashed lines represent the level structure at a five times larger detuning $\Delta$. The solid and dashed lines in (b) depict the same level structure but for a three times larger anharmonicity $\gamma$, showing that our theory is not only applicable to transmons but also to other qudits with larger anharmonicity.
	(c) Using the qudit as a sensor for an unknown drive: The frequency of the first transition $\omega_{01}$ compared to the bare frequency $\omega_{01}^0=\omega_\mathrm{q}-\gamma/2$ and the resulting anharmonicity provide the detuning $\Delta$ and the drive amplitude $A_\mathrm{D}$.}
	\label{fig:analytics}
\end{figure}

The resulting energy levels of Eq.~(\ref{Eq:En}) in the laboratory frame are depicted in Fig.~\ref{fig:analytics} for different $\Delta$ and $\gamma$.
In the case of very large detuning between drive and qubit transition, $\Delta \gg A_\mathrm{D},\gamma$ Eq.~(\ref{eq:alpha_eq}) simplifies to $\alpha \approx A_\mathrm{D}/2\Delta$.
As $\alpha$ is small in this case, we can neglect higher order perturbative terms and we get $E_k /\hbar=\omega_\mathrm{q} k - \frac{\gamma}{4}k(k+1) -\gamma \left|  \frac{A_\mathrm{D}}{2\Delta}  \right|^2 k$, i.e.\@ all transitions shift in parallel and the shift is linear in applied drive power, as expected for the first transition in the dispersive Jaynes-Cummings Hamiltonian.

If we plot the frequency difference between the first and second transition over the change of the first transition for different detunings $\Delta$ and drive amplitudes $A_\mathrm{D}$, we get a plot which can be used to inversely look up the drive amplitude and detuning from the measured values (Fig.~\ref{fig:analytics}~(c)). 
This enables us to take the qudit as a sensor for the amplitude and frequency of arbitrary microwave drives, as demonstrated and further explained in Sec.~\ref{subsec:sensing}. These tones can result from harmonic oscillators like microwave resonators, magnon systems or other microwave circuitry on the chip, where the goal is to study the system properties or the crosstalk, or they are directly fed from a microwave source.
Application potential is seen in sensing of external or internal microwave signals, for example for the characterization of microwave transmission down to the qubit.

\section{\label{sec:experiment}{Experiment}}
We now apply our theoretical findings to an experiment in order to verify our theory. 
We first experimentally study the level transitions of the qudit in the presence of an external microwave drive and demonstrate good agreement of the relative shifts with master equation simulations.
Subsequently, we quantitatively validate the feasibility of this scheme to sense the amplitude and relative frequency of a microwave signal in a wide frequency range.

\subsection{Sample and Setup}
\begin{figure}[tb]
\includegraphics[width=\columnwidth]{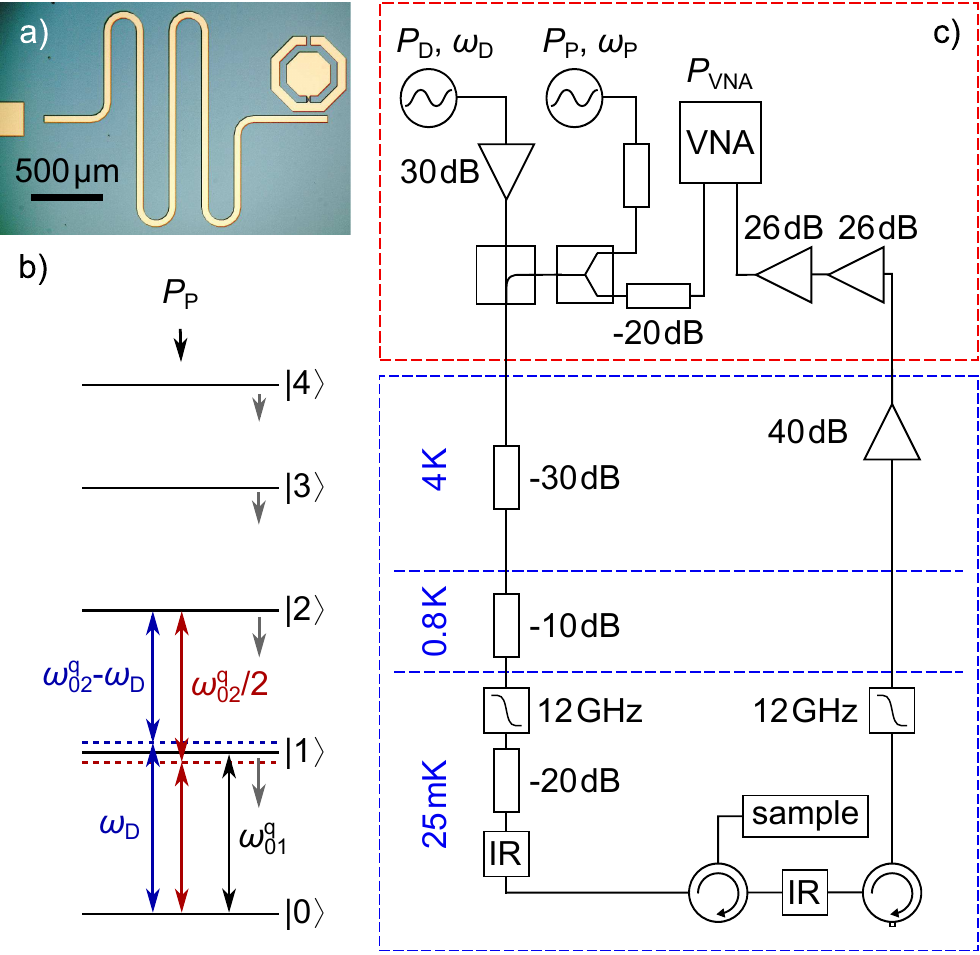}
\caption{(a) Micrograph of the transmon sample. The microwave tones are coupled via the transmission line (left), band-pass filtered by the resonator (center) and finally reach the transmon (right).
(b) Level scheme for the strongly driven circuit. Grey arrows indicate the AC Stark shift of the qudit by the strong drive. Also shown are multi-photon transitions by the drive (red) and by drive and probe combined (blue).
(c) Schematic diagram of the employed microwave setup for reflection measurements. The resonator probe tone $P_\mathrm{VNA}$ and transmon probe tone $P_\mathrm{P}$ are merged in a power combiner and the drive tone $P_\mathrm{D}$ is addedd via a directional coupler. The combined signals are sequentially attenuated at various temperature stages of the cryostat before reaching the the sample. Circulators are placed after the sample to screen thermal noise.
}
\label{Fig2}
\end{figure}

\begin{figure}[tb]
\includegraphics[width=\columnwidth]{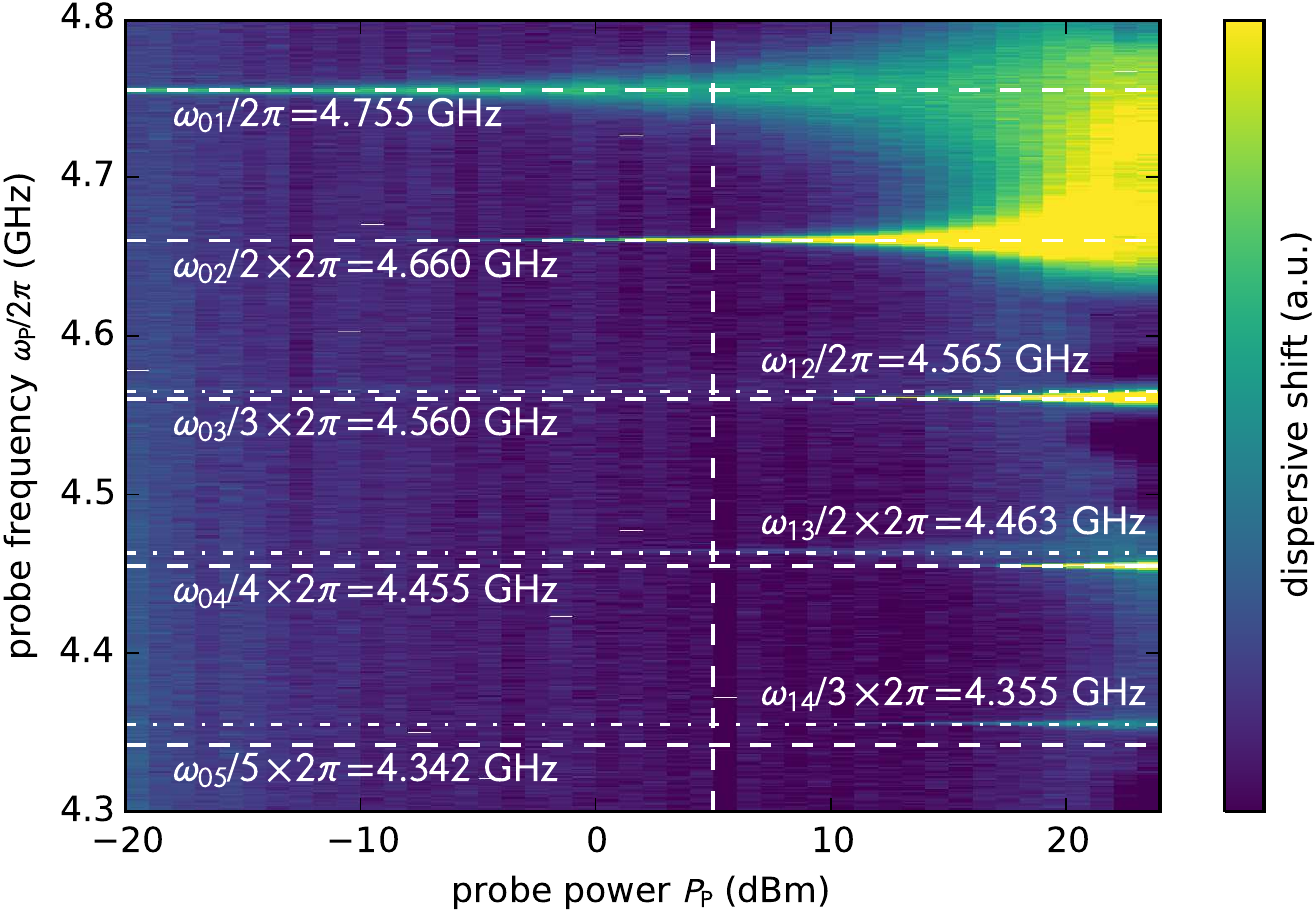}
\caption{Qudit spectroscopy for the concentric transmon at varying probe power and frequency without an additional drive tone. The emergence of multi-photon transitions at higher probe powers is clearly visible. The shadow-like transition next to the $\omega_{04}/4$ transition and the transition around $\SI{4.36}{\giga\hertz}$ can be assigned to transitions between higher states, showing that the system was initially not in the ground state at these high powers. The vertical dashed line at $P_\mathrm{P} = \SI{5}{\decibel m}$ indicates the probe power chosen for the following measurements, as an even higher probe power would strongly increase the linewidth for the first transition and hence reduce the visibility.}
\label{fig:powerspectrum}
\end{figure}

The quantum circuit is a single-junction concentric transmon capacitively coupled to a $\lambda/2$ readout resonator, similar to Ref.~\cite{BraumuellerAPL16}, see Fig.~\ref{Fig2}~(a). 
From microwave spectroscopy, we determine the fundamental qubit frequency $\omega^\mathrm{q}_{01}/2\pi=\SI{4.755}{\giga\hertz}$, being $\Delta_\mathrm{R}/2 \pi=\SI{3.818}{\giga\hertz}$ red-detuned from the resonator mode at $\omega_\mathrm{r}/2\pi = \SI{8.573}{\giga\hertz}$ with a coupling of $g/2\pi = \SI{71.5}{\mega\hertz}$.
The transition frequencies of the anharmonic levels $\omega^\mathrm{q}_{0 \rightarrow n}/n$ are determined from power spectroscopy, shown in Fig.~\ref{fig:powerspectrum}.
From this spectrum, we can determine the charging energy $E_\mathrm{C}/h = \SI{197.7}{\mega\hertz}$ and  the Josephson energy $E_\mathrm{J}/h = \SI{15.5}{\giga\hertz}$.

The chip features a microstrip design, where the capacitance pads are made from low-loss TiN and the junction is an Al/$\mathrm{AlO_x}$/Al structure. 
We measured coherence times up to $T_1 \approx \SI{30}{\micro\second}$ and $T_{2}^\mathrm{Ramsey} \approx \SI{50}{\micro\second}$. 
The resonator is coupled to a $50\,\Omega$ matched microwave transmission line, where all microwave signals are applied, meaning that all tones to the qubit, i.e.\@ $\omega_\mathrm{D}$ and $\omega_\mathrm{P}$, are band-pass filtered by the resonator.
The quantum chip is measured in a dilution refrigerator at a base temperature of about $25\,\mathrm{mK}$, see Fig.~\ref{Fig2} for a schematic measurement setup diagram. The chip is placed inside a copper box and mounted to the mixing stage. 

We experimentally investigate the circuit by applying a continuous microwave probe tone of frequency $\omega_\mathrm{P}$ and power $P_\mathrm{P}$ to excite direct or multi-photon transitions from the ground state to higher levels $\ket{k}$.
The level population $\langle \hat{n}^\mathrm{q}\rangle$ with number operator $\hat{n}^\mathrm{q}=\sum_k k \ket{k}\bra{k}$ of the anharmonic circuit is measured via the readout resonator, where higher level populations in the anharmonic oscillator cause an increasingly larger dispersive shift of the resonance frequency. This non-trivial dependence is in detail analyzed in Ref.~\cite{BraumuellerPRB15}.
The resonator readout is done using a vector network analyzer (VNA) with $\omega_\mathrm{VNA}/2\pi=\SI{8.573}{\giga\hertz}$ and $P_\mathrm{VNA}=\SI{-22}{\deci\bel m}$. Data acquisition and analysis is done via qkit \cite{qkit}.

To measure the AC Stark shift up to values of $A_\mathrm{D} \gg \gamma$, the signal of the driving microwave source was amplified at room temperature by a high-power amplifier. As the amplifier is operated partially beyond its compression point, all drive powers $P_\mathrm{D}$ given in the following refer to measured values after the amplifier.

\subsection{Measuring the AC Stark Shift}
\begin{figure*}[tb]
	\includegraphics[width=\textwidth]{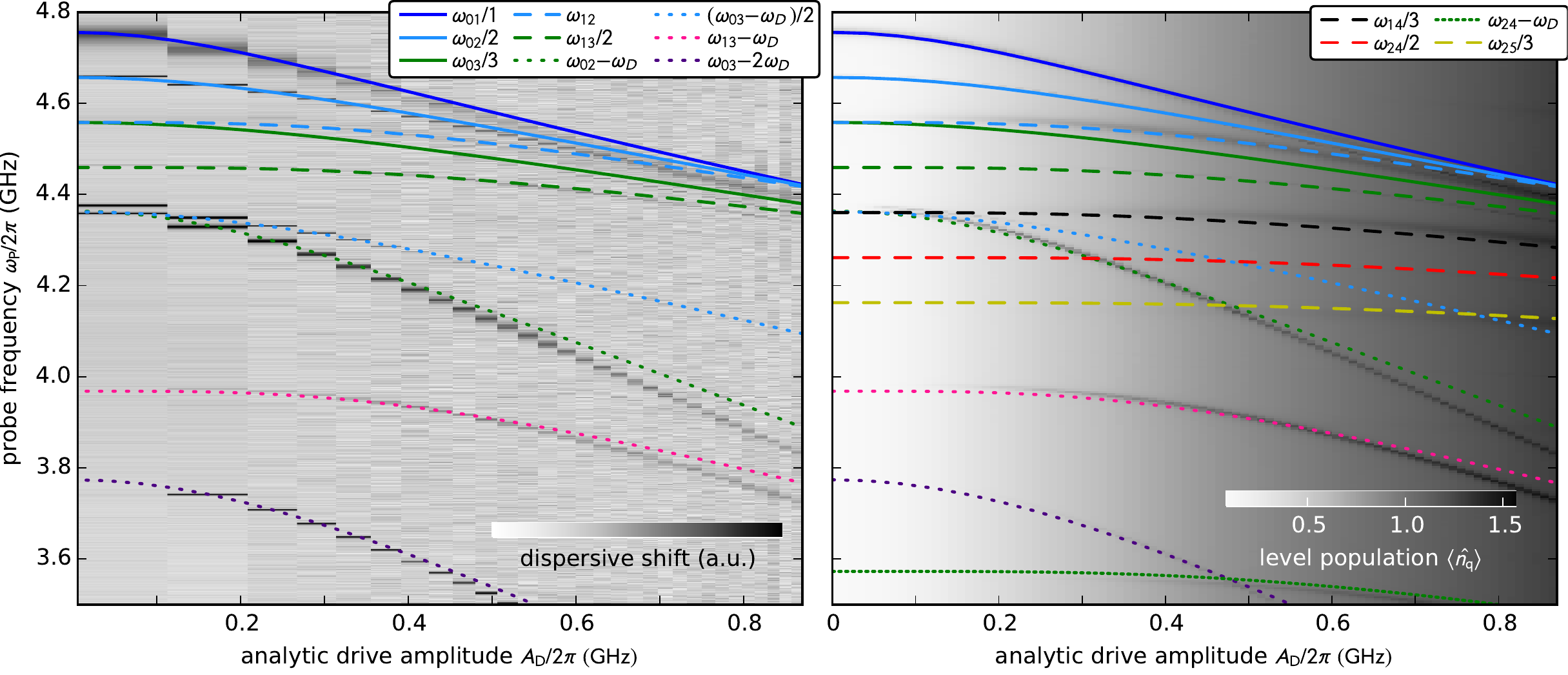}
	\caption{Left: AC Stark shift of the transmon levels under a power-varying drive at $\omega_\mathrm{D}/2\pi = \SI{4.95}{\giga\hertz}$.  The analytic solution from Eq.~(\ref{Eq:En}) (colored lines) fits well to the measured transitions. Additional multi-photon transitions consisting of one or two drive photons and one or two probe photons are shown as dotted lines. Deviations from the analytic solutions are attributed to using a simplified transmon Hamiltonian, Eq.~(\ref{eq:driven_transmon}), which does not represent higher transmon levels correctly.
		The data has been column-wise normalized, as the drive also influences the resonator's resonance frequency.\\
		Right: QuTiP simulations with $\omega_\mathrm{D}/2\pi = \SI{4.95}{\giga\hertz}$. Line colors are the same as in the left plot. Due to a slightly higher $A_\mathrm{P}$, more transitions including the $\ket{4}$ state are visible that can be explained as well with the analytic approximation. }
	\label{fig:shifts}
\end{figure*}

The AC Stark shift as shown in Fig.~\ref{fig:shifts} (left) is spectroscopically determined for fixed probe power $P_{\mathrm{P}}$ and varying drive power $P_{\mathrm{D}}$ between $-6.3$ and $+23.9\,\mathrm{dBm}$ while scanning the probe frequency $\omega_\mathrm{P}$.
Comparing the measured data to the analytic approximation in Eq.~(\ref{Eq:En}), we find a very good agreement for the multi-photon transitions up to the third qudit level.
Additional lines are explained by taking into account that one or more photons can be supplied by the drive (compare Fig.~\ref{Fig2}~(b)).

Comparing the propagating tone with a resonantly driven resonator ($\omega_\mathrm{R}/2\pi = \SI{4.95}{\giga\hertz}$, $g/2\pi = \SI{71.5}{\mega\hertz}$) requires a population of around 40 photons for a shift similar to $A_\mathrm{D}/2\pi = \SI{0.9}{\giga\hertz}$.
In this case, Eq.~(\ref{Eq:jaynescummings}) can not be used for a simple estimation of the shift, as we are no longer in the dispersive regime, where $g/\Delta \ll 1$.
Deviations between the analytic solution and measured data for higher transmon levels are explained by considering a first-order nonlinearity for the qubit Hamiltonian in Eq.~(\ref{eq:driven_transmon}), which is not accurate for higher level transitions.

\subsection{Numerical Simulations}
Numerical simulations are performed with the master-equation solver of QuTiP \citep{qutip1,qutip2} where we use the full transmon Hamiltonian from Ref. \cite{Koch_TransmonPRA07} without a resonator and apply the drive directly to the qudit. 
To restrict the Hilbert space of the simulation, excitations up to the $\ket{9}$ level of the transmon are taken into account, as higher states are above the Josephson potential barrier.
As the applied drives require for a time-dependent solution, we reduce the qudit $T_1, \, T_2$ times to increase the simulation speed.
The basic simulation principle is described in Ref.~\citep{BraumuellerPRB15} and additional information is given in the supplement. The simulation results, shown in Fig.~\ref{fig:shifts}  (right), are in very good agreement with the measured result.
The same deviations from the analytical solution as in Fig.~\ref{fig:shifts} (left) can be found since the transmon model used in the simulation also accounts for higher-order nonlinearity. As the probe power in the simulation is slightly higher than in the experiment, more transitions are visible.

\begin{figure}[tb]
	\includegraphics[width=\columnwidth]{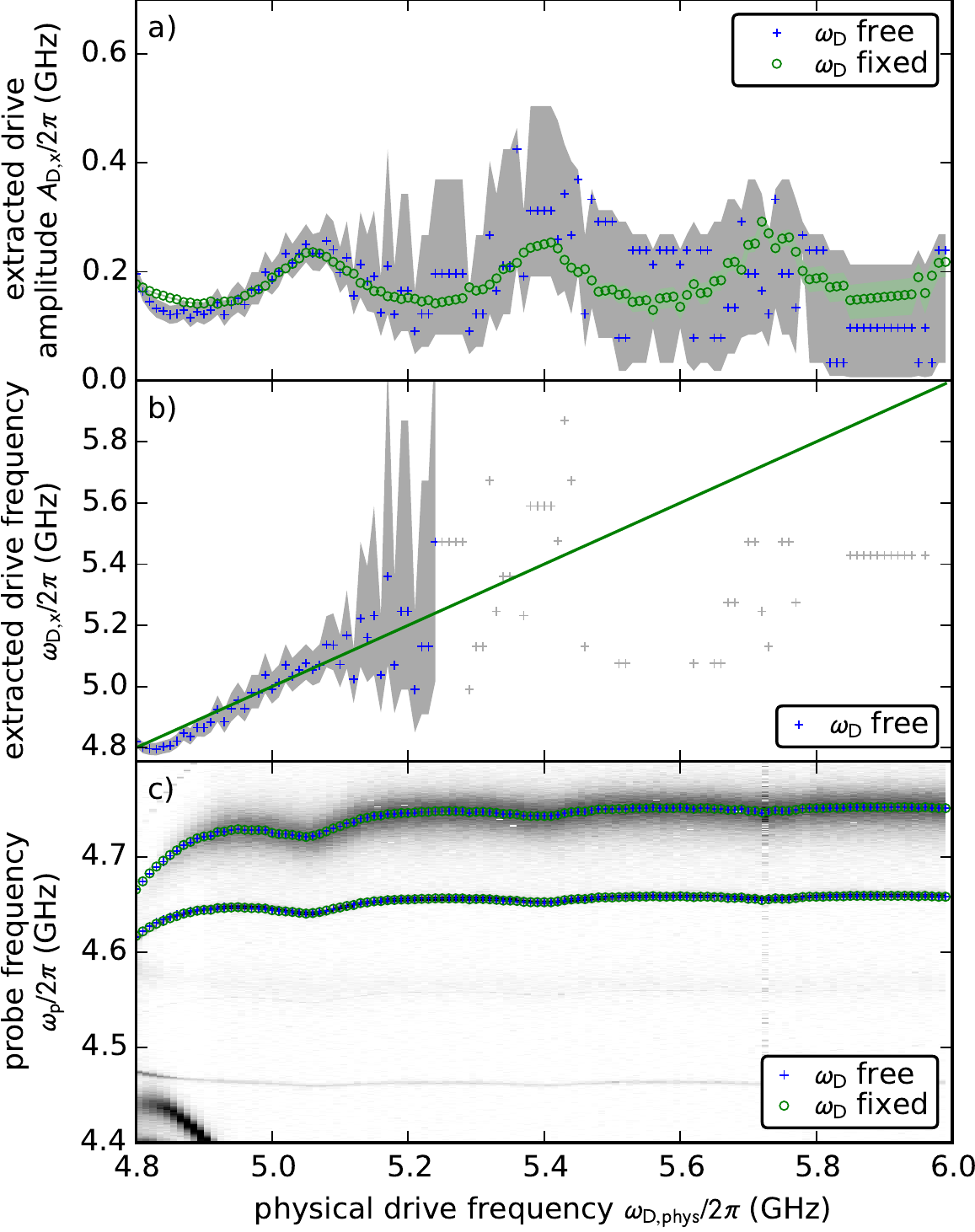}
	\caption{Verification of the sensing technique: Scanning the drive frequency at a constant drive power. From the fitted $\omega_{01}$ and $\omega_{02}/2$ transitions, we calculate the drive amplitude (a) and frequency (b), where both $A_\mathrm{D}$ and  $\omega_{\mathrm{D}}$ are free (blue) or $\omega_{\mathrm{D}}$ is fixed to the known value of the drive frequency (green). The gray and green shaded areas show the uncertainty if we take into account an accuracy of $\pm \SI{1}{\mega\hertz}$ for both $\omega_{01}/2\pi$ and $\omega_{02}/4\pi$. As the frequency determination fails for more than $\SI{0.5}{\giga\hertz}$ detuning between drive and the first qudit transition at $\omega_{\mathrm{D}}/2\pi = \SI{4.755}{\giga\hertz}$, no errorbars are shown for this area.
	To verify the method of fixing $\omega_{\mathrm{D}}$ to the known value, we calculated the  $\omega_{01}$ and $\omega_{02}/2$ transitions from the values in (a) and (b),  plotted them on top of the measurement data (c) and find a good agreement.}
	\label{fig:sensing}
\end{figure}

\subsection{\label{subsec:sensing}{Demonstration of the sensing scheme}}
In order to verify the sensing technique, we fix the drive power and scan the drive frequency in a second experiment. 
Scanning the probe tone, we can identify the qudit $\omega_{01}$ and $\omega_{02}/2$ transitions and then use our analytical model to calculate $A_\mathrm{D}$ and $\omega_{\mathrm{D}}$.
The results are shown in Fig.~\ref{fig:sensing} for $A_\mathrm{D}$ (a) and $\omega_{\mathrm{D}}$ (b) as blue points. To increase the accuracy of the measurement, we can set  $\omega_{\mathrm{D}}$ to the known value of the drive frequency, which greatly improves the precision of the amplitude measurement (green points in Fig.~\ref{fig:sensing}~(a)).
For the mesh shown in Fig.~\ref{fig:analytics}~(c), this means that we select the matching blue line for the known detuning $\Delta = \omega_\mathrm{D} - \omega_{01}$ and search for the right value of $A_\mathrm{D}$.
As the influence of the AC Stark effect on the level transitions decreases with increasing detuning, it is obvious that resolution and accuracy decrease at the same time.

The precision of these measurements is illustrated by the gray and green shaded areas in Fig.~\ref{fig:sensing}~(a) and (b), which visualize the spread of results when varying $\omega_{01}$  and $\omega_{02}/2$ by $\Delta\omega / 2\pi =\pm \SI{1}{\mega\hertz}$, being a worst-case estimate. 
In experiment, this uncertainty originates from using a constant probe power and a step size of $\Delta \omega_{\mathrm{P}}/2\pi = \SI{1}{\mega\hertz}$. 
This results on average in $\Delta A_\mathrm{D} = \pm \SI{92}{\mega \hertz}$, $\Delta \omega_\mathrm{D} = \pm \SI{124}{\mega \hertz}$ for free $\omega_{\mathrm{D}}$ and $\Delta A_\mathrm{D} = \pm \SI{12.5}{\mega \hertz}$ for fixed $\omega_{\mathrm{D}}$. 
For a detuning between drive and first transition of approximately more than $\SI{0.5}{\giga\hertz}$, 
the sensitivity of the level shifts in the frequency detuning of the drive becomes rather poor, see Fig.~\ref{fig:shifts}~(b). 
For this reason we do not show errorbars for
 $\omega_{\mathrm{D}}/2\pi > \SI{5.25}{\giga\hertz}$ and plot the calculated points in gray.
The resolution could potentially be greatly improved by reducing the probe power for the $\omega_{01}$ transition to decrease the transition linewidth and reducing the step size for scanning the probe tone.
With that, the uncertainty would be reduced to $\Delta \omega /2\pi = \pm \SI{100}{\kilo\hertz} $, which results for our measurement on average to $\Delta A_\mathrm{D} = \pm \SI{13.6}{\mega \hertz},\,\Delta \omega_\mathrm{D} = \pm \SI{11.8}{\mega \hertz}$ for free $\omega_{\mathrm{D}}$ and $\Delta A_\mathrm{D} = \pm \SI{1.2}{\mega \hertz}$ for fixed $\omega_{\mathrm{D}}$. 

As an ideal curve shape for $A_\mathrm{D}$  in Fig.~\ref{fig:shifts}~(a), we would expect an increase of  $A_\mathrm{D}$ close to the qudit transition frequencies due to the frequency-dependent dipole moment of the qudit and a monotonous decrease towards a constant value at higher detunings.	
In the measurement however, we see the onset of an increase close to the qudit transition. 
More dominant are oscillations in  $A_\mathrm{D}$ for higher detuning, which can be explained by a non-ideal transmission of the microwave lines in our cryostat. 
This demonstrates the usability of a single non-tunable transmon qudit as a sensor to calibrate the amplitudes, and therefore powers, actually arriving at the qudit position for given input frequency and power, thus making the qudit a local power sensor.
For a given input frequency, the local power coupled to the qudit can be sensed by only spectroscopic measurements, being independent of qubit coherence.
Being able to measure the amplitude of a detuned microwave tone, one can also measure the crosstalk on a quantum chip by exciting or driving a structure at one position on the chip and detect its influence on a qudit at a different location.

\section{\label{sec:conclusion}{Conclusion}}
In summary, we have investigated a superconducting anharmonic multi-level quantum circuit under the influence of a variable classical field drive. 
The AC Stark shift of up to the $3^\mathrm{rd}$ level and multi-photon transitions consisting of probe and drive photons, including virtual energy levels are observed.
The field-induced shift in our measurement is shown to be in good agreement with both a fourth order perturbation theory calculation and a master-equation simulation of the system. 
With our model, we are able to quantify the spurious crosstalk on multi-element quantum chips.
Providing a good amplitude resolution and a local measurement, we can also use the system to calibrate the power at the qudit position and therefore characterize the microwave transmission from room temperature to the actual qudit in a broader frequency range with only spectroscopic measurements.

For the first transition, in the limit of large detuning $\Delta$, the analytic formula is shown to recover the behavior of a Jaynes-Cummings type system, where the shift of the qudit transition frequencies is linear in the population of the resonator, i.e.\@ the drive power. 
In contrast to the Jaynes-Cummings formula, the frequency shift calculated from Eq.~(\ref{Eq:En}) can also be applied in the non-dispersive regime where the detuning $\Delta$ is not large.

We demonstrate the viability of the investigated sensing technique and find uncertainties of tens to hundred MHz for both amplitude and frequency, limited by the measurement resolution for this proof-of-principle demonstration. 
We discussed that these uncertainties could be reduced down to several MHz by an improved measurement scheme. 
Other sensing techniques such as Ramsey pulse sequences for frequency detection or Autler-Townes experiments for power measurement, which offer a higher resolution, are only applicable for small detunings in the order of few MHz.
While the here presented sensing technique offers a lower resolution, it is applicable for a much wider frequency range up to $\SI{\sim 1}{\giga\hertz}$ of detuning. 
With our method it is also possible to measure a signal if both frequency and amplitude are unknown or if the drive source is not controllable, opening the path to a broad set of new measurement applications.

\begin{acknowledgments}
The authors are grateful for quantum circuits provided by D. Pappas, M. Sandberg and M. Vissers.
This work was supported by the European Research Council (ERC) under the Grant Agreement 648011, Deutsche Forschungsgemeinschaft (DFG) within Project No. WE4359/7-1, the Initiative and Networking Fund of the Helmholtz Association, and the state of Baden-W\"urttemberg through bwHPC.
We acknowledges financial support by the Carl-Zeiss-Foundation (A.S.), the Landesgraduiertenf\"orderung (LGF) of the federal state Baden-W\"urttemberg (J.B.) and the Helmholtz International Research School for Teratronics (J.B.).
A.V.U. acknowledges partial support from the Ministry of Education and Science of the Russian Federation in the framework of the contract No. K2-2016-063.
\end{acknowledgments}

\clearpage
\appendix
\include{suppl_content}

\end{document}

%% file: suppl_content.tex
\section{Comparison with resonator population}
\begin{figure}[tb]
\includegraphics[width=0.45\textwidth]{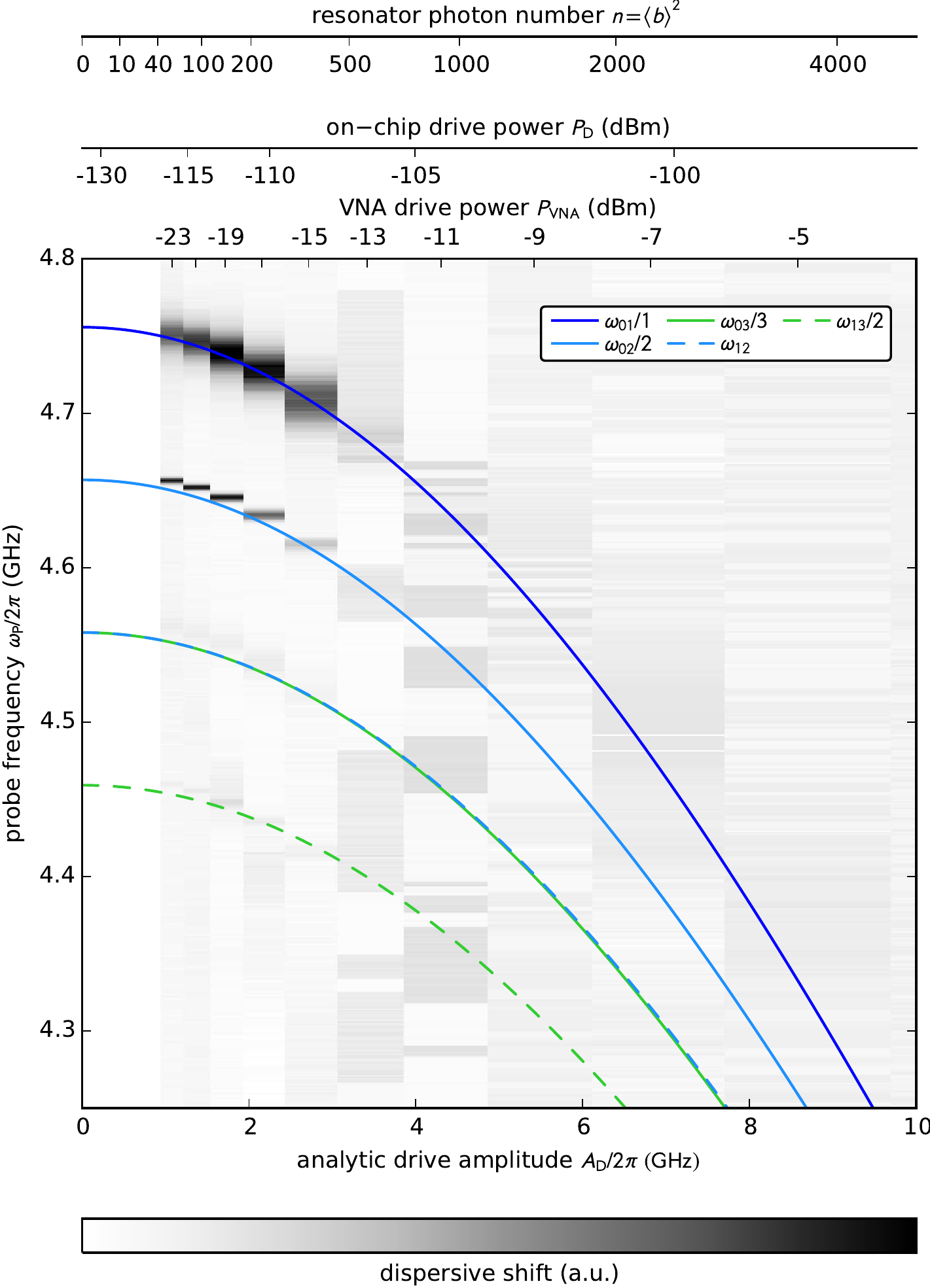}
\caption{AC Stark shift of transmon transition frequencies as obtained through resonator population.
The axes above the image show the average resonator photon population and the effective drive power on the feedline of the chip, both calculated from the analytic drive amplitude $A_\mathrm{D}$ on the bottom of the image. 
The axis on top of the graph shows the room temperature drive power of the VNA probing the resonator. 
A good agreement between our measured data and the analytic solution (colored lines) can be seen. The data is normalized column-wise as the probe frequency of the VNA had to be adjusted for each drive power. The reduced visibility of the qudit for higher drive powers clearly demonstrates the advantages of the off-resonant AC Stark shift.}
\label{Fig:ResonatorDrive}
\end{figure}

In the main part, we report on the AC Stark shift of the qudit levels by applying a strong off-resonant drive tone.
However, the level shift can also be induced by a strongly populated resonator that is coupled to the qudit.
To this end, we experimentally investigated the AC Stark shift of qudit levels when increasing the signal strength used for the probe of the readout resonator and compared measured data to our analytic formula.  See Fig.~\ref{Fig:ResonatorDrive} with different horizontal axes for a comparison of the different power scales.

While the qubit parameters $\omega_\mathrm{q},\,\gamma$ remain unchanged compared to the experiment in the main part, the detuning is now $\Delta/2\pi = ({\omega_\mathrm{r}-\omega_\mathrm{q}+\frac{\gamma}{2}})/{2\pi} =\SI{3.818}{\giga\hertz} $. With a coupling of $g/2\pi = \SI{71.5}{\mega\hertz}$, we calculate the resonator photon number as $n = \langle b \rangle^2 = \frac{A_\mathrm{D}^2}{4 g^2}$. Using the relation 

\begin{equation}
n = 4 P_\mathrm{D} \frac{Q_\mathrm{l}^2}{Q_\mathrm{c} \hbar \omega_\mathrm{r}^2}
\label{eq:app:p_to_n}
\end{equation}
we estimate the driving $P_\mathrm{D}$ power on the feedline. $Q_\mathrm{c}$ is the coupling quality factor of the readout resonator, $Q_\mathrm{l} = \frac{Q_\mathrm{c} Q_\mathrm{i}}{Q_\mathrm{c}+Q_\mathrm{i}}$ the loaded and $Q_\mathrm{i}$ the internal quality factor.
Compared to the output power of the VNA at room temperature, we see a constant attenuation of $\SI{-93}{\deci\bel }$ which can be easily explained by considering  $\SI{-28}{\deci\bel }$ attenuation at room temperature and $\SI{-65}{\deci\bel }$ in the cryostat.

\clearpage

\section{Perturbation theory calculation}
For increasing drive amplitudes, $\alpha$ in Eq.~(10) in the main part gets larger and we have to take more terms up to the fourth order in perturbation theory.
To fourth order, the energy of the states is calculated as
\begin{widetext}
\newcommand{\product}[2]{\bra{#1}\tilde{H}^1_\mathrm{R}\ket{#2}}
\begin{eqnarray}
\label{Eq:4thorder}
\tilde{E}_k &\approx& \tilde{E}_k^0 + \underbrace{\product{k}{k}}_{=0} + \sum_{m\neq k} \frac{\left| \bra{m}\tilde{H}^1_\mathrm{R}\ket{k} \right|^2}{\tilde{E}_k^0 - \tilde{E}_{m}^0}
+ \sum_{m\neq k} \sum_{l\neq k} \frac{\product{k}{m}\product{m}{l}\product{l}{k}}{\left(  {\tilde{E}_k^0 - \tilde{E}_{l}^0}  \right)\left(  {\tilde{E}_k^0 - \tilde{E}_{m}^0}  \right)}
- \sum_{m\neq k}\underbrace{\product{k}{k}}_{=0} \frac{\left|  \product{k}{m}  \right|^2}{\left(  {\tilde{E}_k^0 - \tilde{E}_{m}^0}  \right)^2}\nonumber\\
&&+ \sum_{m\neq k} \sum_{l\neq k} \sum_{p\neq k} \frac{\product{k}{p} \product{p}{m}\product{m}{l}\product{l}{k}}{\left(  {\tilde{E}_k^0 - \tilde{E}_{l}^0}  \right)\left(  {\tilde{E}_k^0 - \tilde{E}_{m}^0}  \right)\left(  {\tilde{E}_k^0 - \tilde{E}_{p}^0}  \right)}
- \sum_{m\neq k} \sum_{l\neq k} \frac{\left|  \product{k}{m}  \right|^2 \left|  \product{k}{l}  \right|^2}{\left(  {\tilde{E}_k^0 - \tilde{E}_{m}^0}  \right)^2 \left(  {\tilde{E}_k^0 - \tilde{E}_{l}^0}  \right)}\nonumber\\
&&+\underbrace{\product{k}{k}}_{=0} \, \left(  \dots  \right)
\end{eqnarray}
\end{widetext}

The full expansion of this term becomes lengthy and would not bring additional insight which is why we do not present it here. For all plots in the main part of the paper, the fourth order solution Eq.~(\ref{Eq:4thorder}) has been used.

\section{Numerical simulation}
The Hamiltonian used in the master equation simulation reads
\begin{eqnarray}
H_\mathrm{sim}/\hbar&=&\sum_k \frac{E_k}{\hbar}•\ket{k}\bra{k} + A_\mathrm{P} (a+a^\dagger)\cos{\omega_\mathrm{P} t}\nonumber\\
&& + A_\mathrm{D} (a+a^\dagger)\cos{\omega_\mathrm{D} t}\,.
\end{eqnarray}
The transmon Hamiltonian is expressed in the most general way as an anharmonic multi-level system in its eigenbasis $\ket{k}\bra{k}$ with eigenenergies $E_k$, following the approaches of Ref.~\cite{Koch_TransmonPRA07}.
The values of $E_k$ are calculated using the solutions for the Mathieu functions. $a$ and $a^\dag$ are the anharmonic creation and annihilation operators, taking into account the different coupling strengths of the transmon levels.

The other terms in the simulation Hamiltonian take into account the applied probe and drive microwave tones with drive strengths $A_\mathrm{P},\,A_\mathrm{D}$.
For the simulation we start with the qudit in the $\ket{0}$ state and compute the time evolution of the Lindblad master equation with the dissipative terms
\begin{eqnarray}
\sqrt{\frac{n_\mathrm{therm}+1}{T_1}}a,\quad \sqrt{\frac{n_\mathrm{therm}}{T_1}}a^\dag,\quad \sqrt{\frac{1}{T_2}}a^\dag a
\end{eqnarray}
where $n_\mathrm{therm}$ is the thermal population of the qudit and $T_1$ and $T_2$ are the qubit relaxation and dephasing times.
We use $n_\mathrm{therm} = 0.1$ to account for the electronic temperature of the chip.

\section{Coherence time and numerical simulation}
\begin{figure}[tb]
\includegraphics[width=0.45\textwidth]{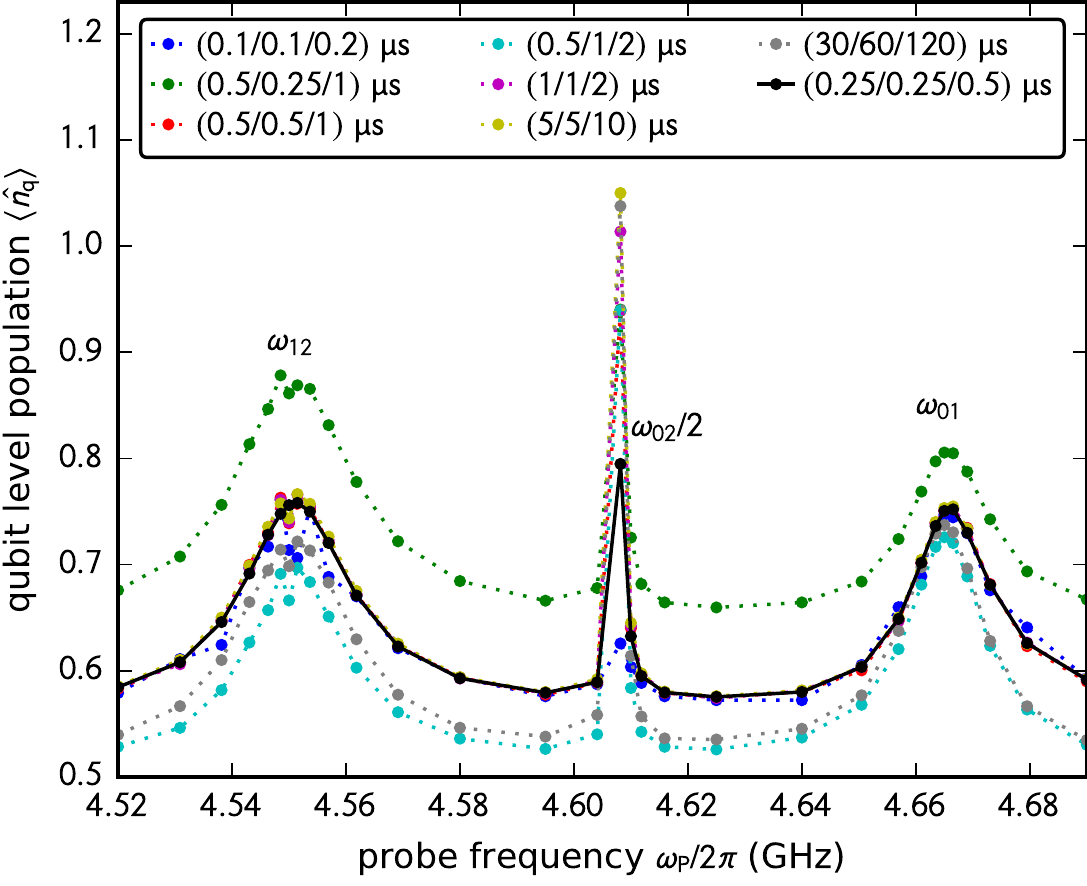}
\caption{Simulation of the first transitions in the spectrum at a drive amplitude of $A_\mathrm{D}/2\pi = 0.3 \,\mathrm{GHz}$ for different $(T_1/T_2/T_\mathrm{sim})$ times. It can be clearly seen that the transition frequencies do not depend on the choice of times. However, the absolute level population varies, as the system is in a driven steady state at the end of the simulation time and therefore the absolute level population depends on the strength of the decay channels.
The combination of  $T_1 = \SI{0.25}{\micro\second},\; T_2 = \SI{0.25}{\micro\second},\; T_\mathrm{sim} = \SI{0.5}{\micro\second}$ was chosen as a good compromise between visibility of all levels and computation time.
The lines are guides for the eye.
\label{Fig5}}
\end{figure}
These numerical solutions are restricted by computation time limitations depending strongly on the circuit's parameters. These quantities have been chosen carefully to minimize the computation times while still allowing for quantum coherence effects. 
The transmon's Hilbert space is restricted to 10 levels, as the $\ket{9}$ state is the highest bound state in the Josephson potential. Including less levels however was seen to reduce the shift of higher level transitions.
Crucial to the computation time is the total simulation length, which can be significantly reduced by reducing the system's $T_1$ and $T_2$ times.
For the simulation we used $T_1={T_2}=\SI{250}{\nano\second}$, being considerably shorter than the experimentally observed values.
The validity of this approach is been verified by performing simulations with different coherence times, see Fig.~\ref{Fig5}, where it can be seen that the AC Stark induced shift does not depend on the chosen coherence times.

During optimization of the simulation parameters, it turned out that computing the time evolution for $T_\mathrm{sim} = 2 \,T_1 $ is sufficient, as the system's dynamics is then dominated by the strong drive and probe tones. To average out the effect of the drive, the level population $\braket{\hat{n}^\mathrm{q}}$ is averaged over a period of $T_1/4$ at the end of the simulation time. The simulations have been performed on the high performance computing cluster \textit{bwUniCluster}.